\documentclass[prb,aps,twocolumn,showpacs,showkeys]{revtex4}

\usepackage{graphicx}
\usepackage{dcolumn}
\usepackage{bm}

\begin{document}
\def \BST{Ba$_{0.5}$Sr$_{0.5}$TiO$_3$ }
\def \er{$\varepsilon_{r}$ }
\def \tand{$tan\delta$ }
\def \BSTA{Ba$_{1-x}$Sr$_x$TiO$_3$ }
\def \AlO{Al$_2$O$_3$ }
\def \x{$\times$ }
\def \k{$\kappa$ }
\def \t{$\theta$ }
\def \e0{$\varepsilon_{r0}$ }
\def \eb{$\varepsilon_{rb}$ }
\def \~{$\approx$ }


\title{Improvement of dielectric loss tangent of doped \BST thin
films for tunable microwave devices}

\author{K. B. Chong$^{a, *}$, L. B. Kong$^{b}$, Linfeng Chen$^{b}$, L. Yan$^{a}$, C. Y. Tan$^{a}$, T. Yang$^{a}$, C.K. Ong$^{a}$ and T. Osipowicz$^{c}$}
\address{$^{a}$Centre for Superconducting and Magnetic Materials, Department of Physics, National University of Singapore, 10 Kent Ridge Road, Singapore 119260 \\ $^{b}$Temasek Laboratories, National University of Singapore,\\10 Kent Ridge Crescent, Singapore, 119260 \\ $^{c}$Research Centre for Nuclear and Microscopy, Department of Physics, National University of Singapore, 10 Kent Ridge Road, Singapore 117602 }
\email{scip1295@nus.edu.sg}

\begin{abstract}
\AlO doped \BST (BST) thin films, with different \AlO contents,
were deposited on (100) LaAlO$_3$ substrate, by pulsed laser
deposition (PLD) technique to develop agile thin films for tunable
microwave device applications.  The dielectric properties of \AlO
doped BST films were determined with non-destructive dual
resonator near 7.7 GHz.  The effects of \AlO doping are the
significant reduction of dielectric constant, dielectric loss and
tunability, compared with pure BST thin film.  The figure of merit
for \AlO doped BST films have also been shown  to increase with
\AlO content. Consequently, \AlO doped BST films have the
potential to be exploited in tunable microwave device
applications.
\end{abstract}

\maketitle

\section{INTRODUCTION}
It is a challenging task for microwave communication researchers
to develop tunable microwave devices.  One of the approaches is to
use a ferroelectric thin film as a buffer layer, while the
dielectric constant \er of the ferroelectric thin film can be
tuned by applying an electric field and hence the working
frequency of the microwave devices can be tuned
accordingly\cite{1,2,3,4}. The forerunner candidate of a hybrid
multi-layer thin film system involves \BSTA (BST) where $x$ is the
concentration of the constituent which can be tuned as well. An
additional advantage of BST is the Curie temperature (T$_C$) of
BST dependence on $x$.  Therefore, the working temperature of the
tunable microwave devices can be tuned as well\cite{5,6,7,8}.
Unfortunately, the dielectric loss \tand of the BST is normally
high, which is around 0.03 for ${x=0.5}$ in the present study on
pulsed laser deposited thin film. The reported values of
$tan\delta$ for BST from different research groups are varied.  In
general, the values of \tand are too high for BST to be useful to
be of practical use in microwave tunable devices.  It is a general
belief that \tand values have to be lower than 0.01 for BST thin
films to be useful in the microwave devices\cite{9,10,11}. In the
present study, we chose to address the problem of dielectric loss
by doping the BST with low loss oxide\cite{12,13,14}.  The \AlO is
chosen for its low microwave dielectric loss as a dopant.  The
dielectric properties of doped ferroelectric film is measured by a
home-made microstrip dual resonator near 7.7 GHz. The result was
then checked by measured the \er of LaAlO$_3$ single crystal
substrate.

\section{EXPERIMENTAL}
BST target with diameter 2.5 cm was prepared using BaTiO$_3$,
SrTiO$_3$ powders with ratio 1:1, via the conventional ceramic
processing.  The BaTiO$_3$ and SrTiO$_3$ powders were mixed and
calcined at 950 $^o$C for 1 hour before they were compacted and
sintered at 1350 $^o$C for 4 hours.  The \AlO doped BST thin films
were deposited on (100) LaAlO$_3$ single crystal substrates with
size of 10 \x 5 \x 0.5 mm$^3$, by PLD with a KrF excimer laser at
repetition rate of 5 Hz, and the average pulse energy was 250 mJ.
BST target with  \AlO plates on its surface was employed in the
film deposition. The \AlO doped BST films deposition as carried
out at substrate temperature of 650 $^o$C and an oxygen pressure
of 0.2 mbar for 45 minutes. The post-deposition annealing was done
in the PLD chamber at the same temperature but with a higher
oxygen pressure of 1.0 atm. The optimum distance between substrate
and target was 4.5 cm for this temperature and pressure. The
content of \AlO in the deposited \AlO doped BST films are
controlled by coverage area of \AlO over the BST target. BST film
without doping was also deposited for comparison.  The films
produced from the targets with 10 percents, 20 percents, 30
percents and 40 percents \AlO coverage of the BST target were
abbreviated to BSTA1, BSTA2, BSTA3 and BSTA4. The \AlO content in
the films were characterized by Rutherford Backscattering Analysis
(RBS), and the relative concentration of \AlO in BSTA1: BSTA2:
BSTA3: BSTA4 were found to be 1 : 4 : 6 : 9 .\\
Structural phase composition and crystallization of the \AlO doped
BST thin films were determined by X-ray diffraction (XRD), using a
Philips PW 1729 type X-ray diffractometer. Surface morphology was
examined by scanning electron microscopy (SEM), using a JEOL
JSM-6340F type field emission scanning electron microscope. The
dielectric properties of \AlO doped BST films in terms of \er and
\tand were measured by a home-made non-destructive microstrip
dual-resonator method at room temperature and microwave frequency
~7.7 GHz\cite{15,16}. The microstrip dual-resonator patterned on a
TMM10i microwave substrate, consists of two planar half-wavelength
resonators coupled through a gap of 36${\mu m}$. The film under
test was placed on top of the microstrip circuit, covering the gap
between two microstrip resonators.  The \er and \tand of the films
were derived from the resonant frequencies $f$$_1$, $f$$_2$ and
the corresponding quality factor $Q$$_1$, $Q$$_2$, of the
microstrip dual-resonator. In the study of the electric field
dependence of the \AlO doped BST thin films, a maximum dc voltage
of 2.1 kV was applied through two electrode pads on the microstrip
circuit board across a gap of about 2.6 mm, corresponding to a
maximum electric field of ~8.1 kV/cm.

\begin{table}
\caption{\label{Tab:1}Dielectric properties of the Al$_2$O$_3$
doped BST films at microwave frequency $\sim$ 7.7 GHz}
\begin{ruledtabular}
\begin{tabular}{lcccc}
\ Thin films & \er                    & \tand        & Tunability    & \k   \\
             &                        & [for 0 bias] & [percentage]  &
             \\  \hline
BST          &1622                    &0.030         &22.0           &7.33          \\
BSTA1        &1387                    &0.021         &19.7           &9.38          \\
BSTA2        &1311                    &0.015         &17.9           &11.93      \\
BSTA3        &950                     &0.0123        &16.4           &13.67      \\
BSTA4        &870                     &0.011         &15.9 &14.45 \\
\end{tabular}
\end{ruledtabular}
\end{table}

\section{RESULTS and DISCUSSION}
Fig. 1 shows the XRD patterns of the deposited thin films. It is
found that all the films have a single phase perovskite structure.
The intensity peak was broadened with increasing \AlO dopant,
which means that the grain size of the \AlO doped BST films
decreased as a result of increased \AlO content. The \AlO was not
detected by XRD due to the relatively small amount of \AlO
compared with BST.  However, for the BSTA4 film, two unidentified
peaks were observed at 2\t \~ 39.39$^o$ and 2\t \~ 43.01$^o$
(marked $\diamondsuit$ as in Fig. 1). These peaks are attributed
to a second phase caused by the excessive \AlO content. The second
phase might be undesirable for the film to be applied in microwave
devices.  This would put a limit on the amount of
dopant.\\
  Fig. 2 displays the surface morphology of the \AlO doped
BST thin films by SEM.  The films exhibited a dense
microstructure, which was greatly modified by \AlO doping. It is
found that the surface roughness increases with the increase of
\AlO content, and it is consistent with the XRD data. Furthermore,
the thicknesses of the thin films were estimated as
500 nm in average from a cross-sectional SEM image.\\
The \er of the \AlO doped BST films as a function of applied
electric field are shown in Fig. 3.  The pure BST films have \er
=1622 that is higher than doped films, BSTA1(\er =1387), BSTA2(\er
=1311), BSTA3(\er =950) and BSTA4(\er =870). The\er of \AlO doped
BST films decrease with increasing \AlO content.  The high \er
values are due to the fact that all thin films are epitaxially
c-axis oriented, which are indicated by (100) and (200) peaks in
XRD patterns.  The highly c-axis oriented films provide a strong
polarization direction compared to randomly oriented samples, and
tends to form a concentrated polarization which results in higher
\er. Fig. 4 depict the plot for \tand of \AlO doped BST films as a
function of the applied electric field.  The \tand decreased as
0.030, 0.021, 0.015, 0.012 and 0.011, corresponding to the samples
BST, BSTA1, BSTA2, BSTA3 and BSTA4, respectively.\\
It is known that tunable microwave device applications, require a
high dielectric constant and low dielectric loss.  The dielectric
tunability was calculated by the formula,
\begin {eqnarray}
tunability&=&\frac{\varepsilon_{r0}-\varepsilon_{rb}}{\varepsilon_{r0}},\label{eq:1}
\end {eqnarray}
where \e0 and \eb represent the dielectric constant value at zero
applied electric field and the maximum applied electric field,
respectively.  The dielectric properties of \AlO doped BST films
were listed in Table 1. Identical to dielectric constant and loss
tangent, the dielectric tunability also reduce as a consequence of
\AlO doping.  We found that the dielectric tunability was reduced
from 22.0 percents (BST thin film) to 15.9 percents (BSTA4). In
tunable microwave devices application, figure of merit \k is
usually used to compare the quality of ferroelectric films, which
is defined as,
\begin {eqnarray}
\kappa&=&\frac{tunability}{tan\delta}.\label{eq:2}
\end {eqnarray}
One can noticed that \k - factors is proportional to \AlO content,
as given in Table 1 in which \k =7.33 (BST) and \k =14.5 (BSTA4).

\section{SUMMARY}
In conclusion, we demonstrated that highly c-axis oriented \AlO
doped BST films could be deposited on (100) LaAlO3 single-crystal
substrate, by PLD technique. It is found that\er, \tand and
dielectric tunability of the \AlO doped BST thin films have
decreased due to increasing \AlO content. As a result of this, \k
- factor were improved significantly. The \AlO substitution into
BST films is observed to significantly modify the microstructure
of the films. Our results showed the dielectric properties of BST
ferroelectric thin films can be readily modified to fit the
requirements of microwave device applications with \AlO
doping.\\

One of the Authors (K. B. Chong) would like to thank National
University of Singapore for the financial assistances.

\newpage
\begin{figure}
\caption{\label{Fig:1} XRD patterns of the \AlO doped BST films on
LaAlO$_3$ substrates, (a) BSTA1, (b) BSTA2, (c) BSTA3 and (d)
BSTA4.}
\end{figure}
\begin{figure}
\caption{\label{Fig:2} Surface morphology of the \AlO doped BST
films on LaAlO$_3$ substrates, (a) BSTA1,  (b) BSTA2, (c) BSTA3
and (d) BSTA4}
\end{figure}
\begin{figure}
\caption{\label{Fig:3} Dielectric constant of the \AlO doped BST
films on LaAlO$_3$ substrates as a function of applied electric
field.}
\end{figure}
\begin{figure}
\caption{\label{Fig:4} Loss tangent of the \AlO doped BST films on
LaAlO$_3$ substrates as a function of applied electric field.}
\end{figure}

\end{document}